\documentclass{JHEP3}
\let\ifpdf\relax
\pdfoutput=1 
\usepackage{amsmath}
\usepackage{cite}
\usepackage{ifpdf}
\usepackage{epsfig,multicol}
\usepackage{bm}

\def\d{\mathrm{d}}

\title{Extracting the Distribution Amplitudes of
the ${\boldsymbol{\rho}}$ meson from the Color Glass Condensate}

\author{J. R. Forshaw \\
School of Physics \& Astronomy, University of Manchester, \\
Oxford Road, Manchester M13 9PL, U.K.\\
\email{jeff.forshaw@manchester.ac.uk} }
\author{R. Sandapen \\
D\'epartement de Physique et d'Astronomie, Universit\'e de Moncton, \\
Moncton, N-B. E1A 3E9, Canada.\\
\email{ruben.sandapen@umoncton.ca} }
\preprint{MAN/HEP/2011/05}
\abstract{
We extract the twist-$2$ and twist-$3$ Distribution
Amplitudes (DAs) of the $\rho$ meson using the HERA data on diffractive $\rho$
photoproduction. We do so using
several Colour Glass Condensate (CGC) inspired and a Regge inspired dipole
models.
We find that our extracted twist-$2$ DA is not much model dependent and is
consistent with QCD Sum Rules and lattice predictions. The extracted twist-$3$
DA is more model dependent but is still consistent with the Sum Rules
prediction.
}
\begin{document}
\section{Introduction}
In  previous papers \cite{Forshaw:2010py,Boer:2011fh}, we extracted the
light-cone
wavefunctions of the $\rho$ meson and the corresponding twist-$2$
Distribution Amplitude (DA) using the precise HERA
data~\cite{Chekanov:2007zr,Collaboration:2009xp} on
diffractive $\rho$ photoproduction. In reference \cite{Forshaw:2010py}, we found
that the extracted
twist-$2$ DA, which is sensitive to the longitudinal light-cone wavefunction,
is broader than the asymptotic form $z(1-z)$ and agrees very well with
QCD Sum Rules predictions. We showed in reference
\cite{Boer:2011fh} that this conclusion is not much model
dependent. 
On the other hand, we also found that the data prefer a transverse wavefunction
with end-point enhancement~\cite{Forshaw:2010py} although the degree of such an
enhancement is model dependent~\cite{Boer:2011fh}.

In this paper, we extend our study in two ways. First, we shall repeat
our fits using a CGC-inspired dipole model with a more realistic description of
saturation
effects than the dipole models we used
previously~\cite{Forshaw:2010py,Boer:2011fh}. Second, we
extend the analysis beyond twist-$2$ by extracting also the
twist-$3$ DA which is sensitive to the scalar part of
the transverse light-cone wavefunction. We shall then compare our
extracted twist-$2$ and $3$ DAs with lattice~\cite{Boyle:2008nj} and QCD Sum
Rules predictions~\cite{Ball:2007zt}.

In references~\cite{Forshaw:2010py,Boer:2011fh}, we fit the data
using models for the dipole cross section which have been previously constrained
by $F_2$ data and which also give a good description of the
diffractive structure function ($F_2^{D(3)}$)
data~\cite{Forshaw:2006np,Marquet:2007nf}. Such models may be called ``forward
models'' since, by the optical theorem, the dipole cross section is equal to
the forward dipole-proton elastic scattering
amplitude. Such models do account for saturation, but in a way independent of
the momentum transfer $t$ or the impact parameter $b$. 

More realistic dipole models take into account the $t$ or $b$-dependence
of saturation effects and have been proposed in
references~\cite{Watt:2007nr,Marquet:2007qa,Kowalski:2006hc}. Here, we shall use
the model by Marquet,
Peschanski and Soyez~\cite{Marquet:2007qa} and refer to it as the t-CGC
model. Given
the limited $Q^2 \in [0,41]~\mbox{GeV}^{2}$ range of the HERA data
on diffractive $\rho$ production to which we fit, we do not consider
DGLAP evolution~\cite{Bartels:2002cj,Kowalski:2006hc}.

\section{Amplitudes and cross-sections}
In the dipole model \cite{Nikolaev:1990ja,Mueller:1994jq}, the 
imaginary part of the amplitude  for diffractive $\rho$ production is given
by~\cite{Marquet:2007qa}:
\begin{equation}
\Im \mbox{m} \mathcal{A}_{\lambda} (s,t;Q^2) =
\sum_{h,\bar{h}} \int \d^2  \mathbf{r} \; \d z \;
\Psi^{\gamma^*,\lambda}_{h,\bar{h}} (r,z;Q^2)
\Psi^{\rho,\lambda}_{h,\bar{h}}(r,z)^{*}
e^{-iz\mathbf{r}.\mathbf{\Delta}} \mathcal{N}(x,\mathbf{r},\mathbf{\Delta}) \;,
\label{non-forward-amplitude}
\end{equation}
where $t=-|\mathbf{\Delta}|^2$. $\Psi^{\gamma^*,\lambda}_{h,\bar{h}}(r,z;Q^2)$
and $\Psi^{\rho,\lambda}_{h,\bar{h}}(r,z)$ denote the light-cone wavefunctions
of the photon and $\rho$-meson respectively while
$\mathcal{N}(x,\mathbf{r},\mathbf{\Delta})$ is the dipole-proton scattering
amplitude.

We choose the dimensionless variable $x$
carrying the
energy dependence of the dipole-proton elastic amplitude
$\mathcal{N}(x,\mathbf{r},\mathbf{\Delta})$ to be a modified Bjorken-$x$: 
\begin{equation}
x = x_{\mathrm{Bj}} \left(1 + \frac{4m_f^2}{Q^2} \right)
\label{energy-variable} 
\end{equation}
with  a phenomenological quark mass of
$m_f=0.14~\mbox{GeV}$. This choice coincides with $x_{\mathrm{Bj}}$ at high
$Q^2$
and replaces the latter at low $Q^2$. We
should point out that the
optimal choice for the dimensionless variable that carries the energy dependence
of the dipole-proton amplitude remains an open and
interesting problem~\cite{Motyka:2008jk}.
For instance, it can be argued that the dipole-proton amplitude should not
depend at all on the photon's virtuality~\cite{Ewerz:2011ph,Forshaw:1999uf}.

Assuming that
$\mathcal{N}(x,\mathbf{r},\mathbf{\Delta})=\mathcal{N}(x,r,\mathbf{
\Delta})$, the angular integration can be carried out giving
\begin{equation}
\Im \mbox{m} \mathcal{A}_{\lambda} (s,t;Q^2) = 2\pi
\sum_{h,\bar{h}}  \int \d r \; \d z \; r J_{0}(rz\Delta) \;
\Psi^{\gamma^*,\lambda}_{h,\bar{h}} (r,z;Q^2)
\Psi^{\rho,\lambda}_{h,\bar{h}}(r,z)^{*}
\mathcal{N}(x,r,\mathbf{\Delta}) \;.
\label{non-forward-amplitude-Bessel}
\end{equation}
Note that by setting $t=0$ in equation \eqref{non-forward-amplitude},
we recover the forward
amplitude used in references~\cite{Forshaw:2010py,Boer:2011fh}, i.e.
\begin{equation}
\left.\Im \mbox{m} \, \mathcal{A}_\lambda(s,t;Q^2)\right|_{t=0} = s \sum_{h,
\bar{h}}
\int \d^2 {\mathbf r} \; \d z \; \Psi^{\gamma,\lambda}_{h, \bar{h}}(r,z;Q^2)
\hat{\sigma}(x,r) \Psi^{\rho,\lambda}_{h, \bar{h}}(r,z)^*
\label{forward-amplitude}
\end{equation}
where we have introduced the dipole cross-section
\begin{equation}
\hat{\sigma}(x,r)\equiv \frac{\mathcal{N}(x,r,\mathbf{0})}{s}=\frac{1}{s}\int
\d^2 \mathbf{b} \; \tilde{\mathcal{N}}(x,r,\mathbf{b}) \;.
\label{dipole-xsec-optical}
\end{equation}
This means that the dipole cross-section is independent of the
impact parameter\footnote{The dipole-proton amplitude $\mathcal{N}$ and dipole
cross-section $\hat{\sigma}$ are sometimes used synonymously in the literature
and one
speaks of the $b$-dependence of the dipole
cross-section~\cite{Kowalski:2006hc}.} and that any modelling of saturation at
the level of this cross-section is necessarily $t$-independent. Nevertheless,
the dipole cross-section can be extracted using the high quality $F_2$
data  and can then be used
to predict the imaginary part of the forward amplitude 
for diffractive $\rho$ production and thus the forward differential
cross-section: 
\begin{equation}
\left. \frac{\d\sigma_{\lambda}}{\d t} \right|_{t=0} =\frac
{1}{16\pi} (\Im\mathrm{m} \mathcal{A}_\lambda(s,0))^2 \; (1 +
\beta_\lambda(0)^2)~,
\label{gammap-xsec}
\end{equation}
where $\beta_\lambda(0)$ is the ratio of real to imaginary parts of the forward
amplitude and is computed as in reference~\cite{Forshaw:2010py}. The
$t$-dependence
can be restored as suggested by experiment~\cite{Chekanov:2007zr}:
\begin{equation}
\frac{d\sigma_{\lambda}}{dt}= \left. \frac{d\sigma_{\lambda}}{dt} \right|_{t=0}
\times
\exp(-B|t|)
\label{exponential-t}
\end{equation}
where
\begin{equation}
B=N\left(
  14.0 \left(\frac{1~\mathrm{GeV}^2}{Q^2 + M_{\rho}^2}\right)^{0.2}+1\right)
\label{Bslope}
\end{equation}
with $N=0.55$ GeV$^{-2}$. Note that in doing this, we are assuming that it is a
reasonable approximation to consider the ratio of real to imaginary part of the
the amplitude to be independent of $t$, i.e.
$\beta_{\lambda}(0)\approx \beta_{\lambda}(t)$.

In the non-forward case, 
\begin{equation}
\frac{d\sigma_{\lambda}}{dt}  = \frac
{1}{16\pi} (\Im\mathrm{m} \mathcal{A}_\lambda(s,t))^2 \; (1 +
\beta_\lambda(t)^2)~,
\label{diff-xsec}
\end{equation}
where $\beta_\lambda (t)$ is now the ratio of real to imaginary parts of the
non-forward amplitude and is computed in the same way as $\beta_{\lambda}(0)$. 
This
differential cross-section is measured at
HERA~\cite{Chekanov:2007zr,Collaboration:2009xp}. 

The total cross-section, which is
also measured at HERA, is obtained by integrating either
\eqref{exponential-t} or equation \eqref{diff-xsec} over $t$. In the former 
case, this integration is trivially performed analytically for each polarisation
and 
the total cross-section is taken to be $\sigma=\sigma_L + \epsilon
\sigma_T$.\footnote{To compare with the HERA data, we take
$\epsilon=0.98$.} In the non-forward case, we integrate equation
\eqref{diff-xsec} numerically over $t$ to obtain the total cross-section.

\section{Dipole models}

In principle, the dipole-proton scattering amplitude
$\tilde{\mathcal{N}}(x,r,b)$ can be
obtained by solving the Balitsky-Kovchegov (BK)
equation~\cite{Balitsky:1995ub,Kovchegov:1999yj,
Kovchegov:1999ua} which itself can be derived within 
the Colour Glass Condensate (CGC)
formalism~\cite{JalilianMarian:1997jx,JalilianMarian:1997gr,Iancu:2000hn,
Iancu:2001ad,Weigert:2000gi}. However, work is still in progress to implement
in a satisfactory way the impact-parameter dependence in the BK
equation~\cite{Berger:2010sh}. On the other hand, an approximate 
solution for the forward dipole-proton amplitude has been proposed several years
ago by Iancu, Itakura and Munier~\cite{Iancu:2003ge}:
\begin{eqnarray}
\mathcal{N}(rQ_s,x,0) &=&  {\cal N}_0 \left( \frac{r Q_s}{2} 
\right)^{2\left[\gamma_s + \frac{\ln(2/rQ_s)}{\kappa \lambda \ln(1/x)}\right]} 
\hspace*{1cm} \mathrm{for} \hspace*{1cm} rQ_s \le 2 \nonumber \\
&=&  \{1 - \exp[-a \ln^2(brQ_s)]\} \hspace*{1cm} \mathrm{for} 
\hspace*{1cm} rQ_s > 2~,
\label{cgc-dipole}  
\end{eqnarray} 
where the saturation scale $Q_s \equiv (x_0/x)^{\lambda/2}$ GeV. The dipole
cross-section is then given by $\hat{\sigma}(rQ_s,x)=\sigma_0
\mathcal{N}(rQ_s,x,0)$. In reference~\cite{Iancu:2003ge}, the anomalous
dimension $\gamma_s$ was fixed to its LO
BFKL value of $0.63$ while the free parameters $\sigma_0$, $x_0$
and $\lambda$ are determined by fitting to the $F_2$ data, without a charm
contribution. When charm was included in reference~\cite{Kowalski:2006hc}, it
was found that the
saturation
scale decreases dramatically. We used this fit in
reference~\cite{Boer:2011fh} where we referred to it as CGC[$0.63$]. However,
Soyez later showed
that letting the
anomalous dimension $\gamma_s$ vary freely also gives a good $F_2$ fit with no
significant decrease in the saturation
scale~\cite{Soyez:2007kg}. The resulting value of $\gamma_s$ is $0.74$,
close to the value obtained with the RG-improved  NLO BFKL kernel
~\cite{Soyez:2007kg}. Kowalski and Watt confirmed the Soyez fit
in reference~\cite{Watt:2007nr} and we used their fitted parameters in
reference~\cite{Boer:2011fh} where we referred to it as the CGC[$0.74$]
fit. This model has been extended by Marquet, Peschanski and Soyez to include a
$t$-dependence in the saturation scale resulting in what we now refer to as
the t-CGC  model. In the t-CGC model~\cite{Marquet:2007qa}, 
\begin{equation}
 \mathcal{N}(rQ_s(t),\mathbf{r},t)= \mathcal{N}(rQ_s(t),\mathbf{r},0) \times
\exp(-B|t|)
\end{equation}
where
\begin{equation}
 Q_s(t)= (x_0/x)^{\lambda/2}(1 + c\sqrt{|t|}) \;.
\label{sat-scale-tCGC}
\end{equation}
The t-CGC model has two additional free
parameters ($c$ and $B$) compared to its forward counterpart, i.e. CGC[$0.74$].
In reference~\cite{Marquet:2007qa}, these parameters were fixed by a global fit
to a subset of the vector meson
($\rho$,$\phi$ and $J/\Psi$) production data including the earlier $\rho$ data
from
H1~\cite{Adloff:1999kg}. The authors of reference~\cite{Marquet:2007qa}
selected those data that were were not too
sensitive to the precise shape of the meson wavefunction
 and did not include the $\sigma_L/\sigma_T$ data.  
As shown in table \ref{CGCparams}, all other
parameters, including the anomalous dimension $\gamma_s$, have the same values
as
in the forward case. This ensures that the quality
of the $F_2$ fit is unaltered in the t-CGC model compared to that with the
CGC[$0.74$] model.

The CGC models are based on perturbative QCD and, for
that reason, the original fits to $F_2$ were in the range $Q^2 \in
[0.25,45]~\mbox{GeV}^2$, thereby excluding the available data in the low $Q^2$
range, $Q^2 \in [0.045,0.20]~\mbox{GeV}^2$, a kinematic region where
perturbation theory
is not reliable~\cite{Watt:2007nr}. 
In these fits the light quark mass was neglected compared to $Q^2$ so
that the dipole-proton amplitude is evaluated at
$x=x_{\mathrm{Bj}}$ when computing the light quark contribution to the structure
function~\cite{Watt:2007nr}. By using equation \eqref{energy-variable}, we
shall dare 
to extrapolate the use of the CGC dipole models to low $Q^2$, including
$Q^2=0$. In order to legitimate such a
procedure, we verify that these models still describe
the $F_2$ data in the extended range $Q^2 \in [0.045,45]~\mbox{GeV}^2$, where
the numerical difference between $x$ given by \eqref{energy-variable} and
$x_{\mathrm{Bj}}$ can be significant. The
resulting $\chi^2/\mbox{d.p}$ values\footnote{Hereafter $\chi^2/\mbox{d.p}$
stands for $\chi^2/\mbox{data points}$.} thus obtained are shown in
the last column of table                              
\ref{CGCparams}. They indicate that it is reasonable to extrapolate  the use of
the CGC models to low
$Q^2$,
together with the modified Bjorken-$x$ given by equation
\eqref{energy-variable}. 

Finally,  
we shall also use the Regge-inspired FSSat dipole 
model~\cite{Forshaw:2004vv}, for which the original $F_2$ fits were 
inclusive of the low $Q^2$ data, i.e. with $Q^2 \in [0.045,45]~\mbox{GeV}^2$.
This model falls in the category of forward models, like CGC[$0.63$]
and CGC[$0.74$], where saturation effects are assumed to be independent of
momemtum transfer. 

Note that we are using here simple phenomenological parametrizations for 
the dipole-proton scattering amplitude subject to the requirement that they
fit the  $F_2$ data. Such simple parametrizations are convenient for
extracting the meson light-cone wavefunction. We do not consider here the
more sophisticated solutions to the BK equation with
running coupling~\cite{Albacete:2007yr,Albacete:2009fh} or the dipole cascade
model~\cite{Flensburg:2008ag}.

\begin{table}[h]
\begin{center}
\textbf{CGC $F_2$ fits}
\[
\begin{array} 
[c]{|c|c|c|c|c|c|c|c|c|}\hline
\mbox{Model} & \gamma_s & x_0 & \lambda & \sigma_0[\mbox{mb}]
&B~\mbox{GeV}^{-2}&c~\mbox{GeV}^{-1}
&\chi^2/133~\mbox{d.p}&\chi^2/156~\mbox{d.p}\\ \hline
\mbox{CGC[$0.63$]} & 0.63~(\mbox{fixed}) & 1.09 \times 10^{-7} & 0.170 &37.4
&-&-& 119/133 & 135/156\\ \hline
\mbox{CGC[$0.74$]} & 0.74 & 1.63 \times 10^{-5} & 0.216
&27.4&-&-&113/133&145/156\\ \hline
\mbox{t-CGC}&0.74&1.63\times10^{-5}
&0.216&27.4&3.754 &4.077&113/133&145/156\\ \hline
\end{array}
\]
\end{center}
\caption{The fitted parameters of the CGC dipole models obtained from fits to
$F_2$
with $Q^2 \in [0.25,45]~\mbox{GeV}^2$ and $x_{\mathrm{Bj}} \le 0.01$. For each
model, the first $\chi^2/\mbox{d.p}$ value is obtained when the model
is fitted to $F_2$ data with $Q^2 \in [0.25,45]~\mbox{GeV}^2$.
The second $\chi^2/\mbox{d.p}$ value
is a prediction of the fitted model when tested in the extended range $Q^2
\in
[0.045,45]~\mbox{GeV}^2$. 
}
\label{CGCparams}
\end{table}

All the forward models we consider also give a good description of the
diffractive
structure function ($F_2^{D(3)}$) data~\cite{Marquet:2007nf,Forshaw:2006np}.
The non forward models have been applied to describe a number of
processes~\cite{Machado:2008tp,Goncalves:2009za} including (in the approach of
reference~\cite{Watt:2007nr}) inclusive hadron production, for which the first
LHC data are now
available~\cite{Levin:2010dw,Tribedy:2010ab}.

\section{Light-cone wavefunctions}
By analogy with the photon's light-cone wavefunction, the $\rho$ meson
wavefunction can be factorised into a spinor and scalar
part~\cite{Forshaw:2003ki}. In momentum space, 

\begin{equation}
\Psi^{\rho,\lambda}_{h,\bar{h}}(z,\mathbf{k})=\sqrt{\frac{N_c}{4\pi}}
S_{h,\bar{h}}^{\rho,\lambda}(z,\mathbf{k}) \phi_{\lambda}(z,\mathbf{k})  
\label{spinor-scalar-k}
\end{equation}
where 
\begin{equation}
 S_{h,\bar{h}}^{\rho,\lambda}(z,\mathbf{k})=\frac{\bar{u}_{h} (zP^+ ,
-\mathbf{k} )}{\sqrt{z}} e^{\lambda}.\gamma
\frac{v_{\bar{h}}((1-z)P^+,\mathbf{k})}{\sqrt{(1-z)}} 
\label{Slambda}
\end{equation}
where $P^+$ is the ``plus'' component of the $4$-momentum of the meson given by
\begin{equation}
P^{\mu}=\left(P^{+},\frac{M_{\rho}^2}{2P^{+}},
0_{\perp} \right) \;.
\end{equation}
Using the longitudinal polarisation vector
\begin{equation}
e^{L}=\left(\frac{P^{+}}{M_{\rho}},-\frac{M_{\rho}}{2P^{+}},
0_{\perp} \right) 
\end{equation}
we can write
\begin{equation}
S_{h,\bar{h}}^{\rho,L}(z,\mathbf{k})=[S_{h,\bar{h}}^{\rho,L}(z,
\mathbf{k})]_+
+ [S_{h,\bar{h}}^{\rho,L}(z,\mathbf{k})]_-
\end{equation}
where~\cite{Lepage:1980fj} 
\begin{equation}
[S_{h,\bar{h}}^{\rho,L}(z,\mathbf{k})]_+=-\frac{M_{\rho}}{2P^+}
\left[\frac{\bar{u}_h}{\sqrt{z}} \gamma^+
\frac{v_{\bar{h}}}{\sqrt{1-z}} \right]=-M_{\rho} \delta_{h,-\bar{h}}
\label{SL+}
\end{equation}
and 
\begin{equation}
[S_{h,\bar{h}}^{\rho,L}(z,\mathbf{k})]_-=\frac{P^+}{M_{\rho}}
\left[\frac{\bar{u}_h}{\sqrt{z}} \gamma^-
\frac{v_{\bar{h}}}{\sqrt{(1-z)}}\right]=-\frac{m_{f}^{2}+ \mathbf{k}^{2}
}{z(1-z)M_{\rho}} \delta_{h,-\bar{h}} 
\label{SL-}
\end{equation}
so that
\begin{equation}
S_{h,\bar{h}}^{\rho,L}(z,\mathbf{k})=-\frac{1}{M_{\rho}
z(1-z)
}
\left[z(1-z)M^{2}_{\rho} +  m_{f}^{2} +  \mathbf{k}^{2} \right]
\delta_{h,-\bar{h}}\;.
\label{SL}
\end{equation}
To evaluate the transverse spinor wavefunction, we use the polarisation vector
\begin{equation}
e^{T}=\frac{1}{\sqrt{2}}\left(0,0, \mathbf{e_{\perp}^{\pm}} \right)
~\hspace{1cm}~\mbox{with}\hspace{1cm} \mathbf{e_{\perp}^{\pm}}=(1,\pm i) 
\end{equation}
to obtain~\cite{Lepage:1980fj}
\begin{equation}
S_{h,\bar{h}}^{\rho,T(\pm)}(z,\mathbf{k})= \pm
\frac{\sqrt{2}}{z(1-z)} \{ [z \delta_{h\pm,\bar{h}\mp} - (1-z)
]\delta_{h\mp,\bar{h}\pm} k e^{\pm i \theta_k} + m_{f}
\delta_{h\pm,\bar{h}\pm} \} 
\label{ST}
\end{equation}
where we have used the polar representation of the transverse
momentum, i.e. $\mathbf{k}=k e^{i \theta_k}$.

Taking the two-dimensional Fourier transforms of equation
\eqref{spinor-scalar-k}, we obtain the  $r$-space light-cone
wavefunctions~\cite{Forshaw:2003ki}:
\begin{equation}
\Psi^{\rho,L}_{h,\bar{h}}(r,z) = -\sqrt{\frac{N_{c}}{4\pi}}
\delta_{h,-\bar{h}}\frac{1}{M_{\rho}z(1-z)} 
[z(1-z)M^{2}_{\rho} + m_{f}^{2} -  \nabla_{r}^{2}] \phi_L(r,z) 
\label{nnpz_L}
\end{equation}
where $\nabla_r^2 \equiv \frac{1}{r} \partial_r + \partial^2_r$, and 
\begin{equation}
\Psi^{\rho,T(\gamma=\pm)}_{h,\bar{h}}(r, z) = \pm
\sqrt{\frac{N_{c}}{4\pi}}\frac{\sqrt{2}}{z(1-z)}[i e^{\pm i\theta_{r}} 
( z \delta_{h\pm,\bar{h}\mp} - (1-z) \delta_{h\mp,\bar{h}\pm}) 
\partial_{r}+ m_{f}\delta_{h\pm,\bar{h}\pm}] \phi_T(r, z).
\label{nnpz_T}
\end{equation}
Previous work~\cite{Forshaw:2003ki,Marquet:2007qa,Watt:2007nr,Flensburg:2008ag}
has argued that a reasonable assumption for the
scalar part of the light-cone wavefunction for the $\rho$ is of the form
\begin{eqnarray}
\phi^{{\mathrm{BG}}}_\lambda(r,z) &=&
\mathcal{N}_\lambda \;  4[z(1-z)]^{b_{\lambda}} \sqrt{2\pi R_{\lambda}^{2}} \;
\exp \left(\frac{m_f^{2}R_{\lambda}^{2}}{2}\right)
\exp \left(-\frac{m_f^{2}R_{\lambda}^{2}}{8[z(1-z)]^{b_{\lambda}}}\right) \\
\nonumber
& &\times \exp \left(-\frac{2[z(1-z)]^{b_\lambda}
r^{2}}{R_{\lambda}^{2}}\right) \;,
\label{boosted-gaussian} 
\end{eqnarray}
which is referred to as the ``Boosted Gaussian'' (BG)
wavefunction.\footnote{This wavefunction is a
simplified version of that proposed originally
by Nemchik, Nikolaev, Predazzi and Zakharov~\cite{Nemchik:1996cw}.}  
These
light-cone wavefunctions are normalised:
\begin{equation}
\int \d^2 {\mathbf{r}} \d z |\Psi^{\rho,\lambda}(r, z)|^{2} = 1 \;,
\label{normalisation}
\end{equation}
where we have defined
\begin{equation}
|\Psi^{\rho,\lambda}(r, z)|^{2} \equiv \sum_{h,\bar{h}}
|\Psi^{\rho,\lambda}_{h,\bar{h}}(r, z)|^{2} \;.
\label{Psisq-summed}
\end{equation}
This normalisation condition embodies the assumption that the $\rho$ 
consists of the leading $q\bar{q}$ Fock state
only. In addition,
the longitudinal light-cone wavefunction is subject to the leptonic decay width
constraint~\cite{Forshaw:2003ki}:
\begin{equation}
\sqrt{2} f_\rho M_\rho = \frac{N_c}{\pi}  \int_0^1 \d z 
\left.[z(1-z)M^{2}_{\rho} + m_{f}^{2}-\nabla_{r}^{2}]
\frac{\phi_L(r,z)}{z(1-z)}
\right|_{r=0} \;,
\label{longdecay}
\end{equation}
where the decay constant $f_\rho$ is simply related to the experimentally
measured electronic decay width $\Gamma_{\rho \rightarrow e^+ e^-}$ of the
$\rho$~\cite{Forshaw:2010py}.

\section{Fitting the HERA data}

In the
original BG wavefunction, $b_L=b_T=1$ and $R_L^2=R_T^2=12.9~\mbox{GeV}^{-2}$ so
that
the leptonic decay width and the normalisation constraints are satisfied.
However, when this BG
wavefunction is used in conjunction with either the FSSat or any of the CGC
models, none of them
is able to give a good quantitative agreement with the current HERA data. This
situation is considerably improved by
letting $R_\lambda$ and $b_\lambda$ vary
freely and this enhances end-point contributions. We also investigate the
requirement for additional end-point enhancement
in the
transverse wavefunction by using a
scalar wavefunction of the form
\begin{equation}
\phi_T (r,z)= \phi^{{\mathrm{BG}}}_T (r,z) \times [1+ c_{T}
\xi^2 + d_{T} \xi^4]
\label{EG} 
\end{equation}
where $\xi=2z-1$.

We fit to the same data set as in reference \cite{Forshaw:2010py}, i.e. to
total cross-section data, the
ratio of longitudinal to transverse cross-section data and the
decay constant datum for the longitudinally polarised meson, i.e. to a total of
$76$ data points. For the
non-forward models, we also include
the differential cross-section data with $|t| \le 0.5~\mbox{GeV}^2$ ($46$ data
points) resulting in a total of $122$ data points. As in
reference \cite{Forshaw:2010py}, we rescale the overall normalisation of
the data down by $5\%$ which is consistent with the experimental
uncertainty. 

Note that for the non-forward t-CGC model, we make use of the
current data on the differential cross-section to refit
the parameters controlling the $t$-dependence. This means that there are 
$2$ additional free parameters, $B$ and $c$, compared to
the forward FSSat, CGC[$0.63$] and CGC[$0.74$] fits.

Our best fit parameters and resulting $\chi^2/\mbox{d.o.f}$ are given in table
\ref{tab:best-fit-params}. For the t-CGC model, we find two successful fits
which we refer to as t-CGC and t-CGC(alt.).
We choose to record the alternative t-CGC fit because we extract from
it a qualitatively
different shape for the transverse wavefunction: the parameters $c_T$ and $d_T$
go negative in this fit.

We note that our fitted values for the parameters
controlling the $t$-dependence are $B=3.159(3.229)~\mbox{GeV}^{-2}$ and
$c=1.146(1.262)~\mbox{GeV}^{-2}$, where the number in brackets corresponds to
the
alternative t-CGC fit. The values for
$c$ are lower than that obtained in reference ~\cite{Marquet:2007qa} and
which we quote in table \ref{CGCparams}. 

\begin{figure}
\centering
\includegraphics*[width=15.cm]{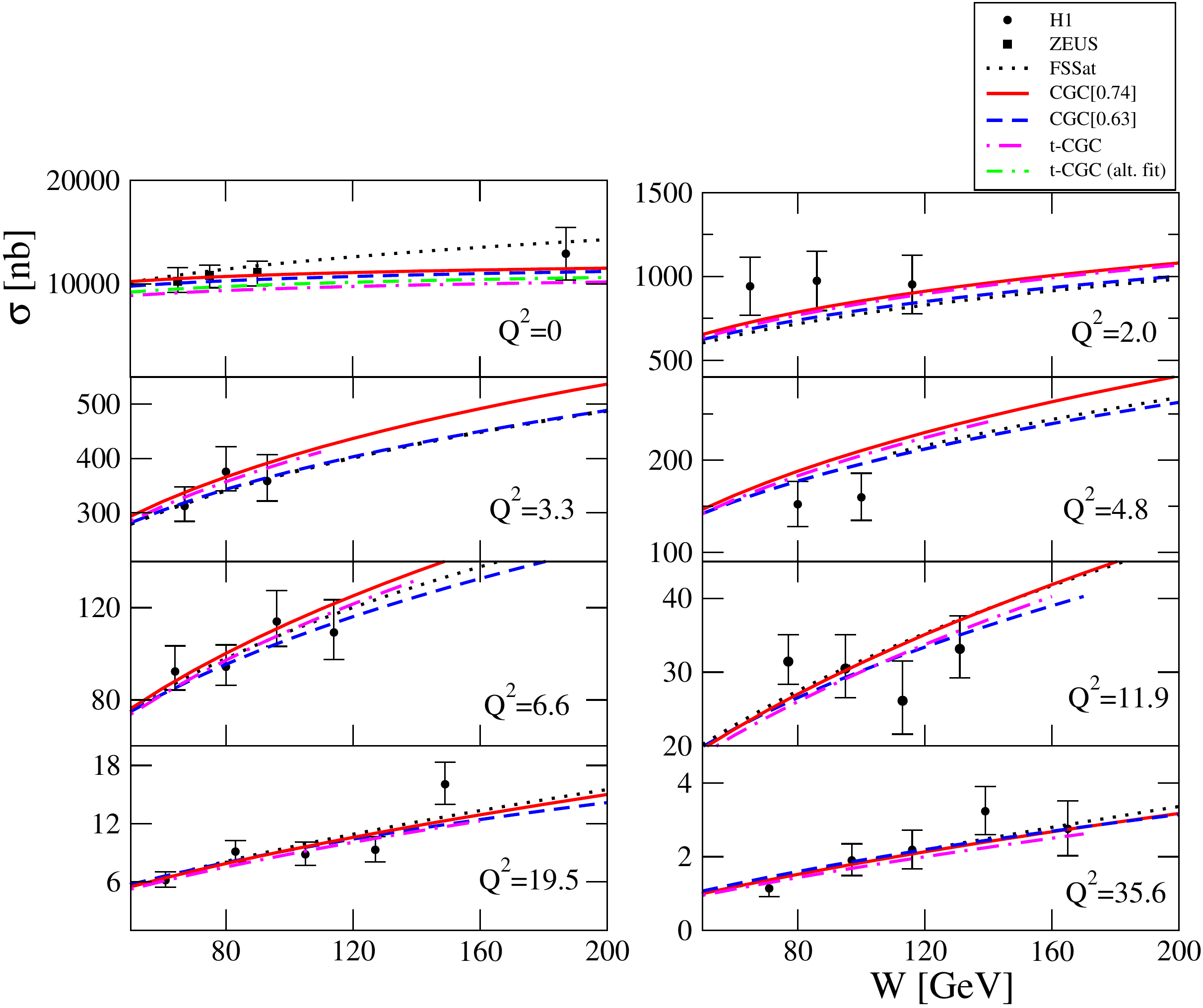}
\caption{Fits to the HERA cross-section data. Solid: CGC[0.74], dotted:
FSSat, dashed: CGC[0.63], dot-dashed: t-CGC. The alternative t-CGC(alt.) fit is
only
shown at $Q^2=0~\mbox{GeV}^{2}$ since at higher $Q^2$ it is indistinguishable
from the t-CGC fit.}
\label{fig:H1-xsec}
\end{figure}

\begin{figure}\centering
\includegraphics*[width=15.cm]{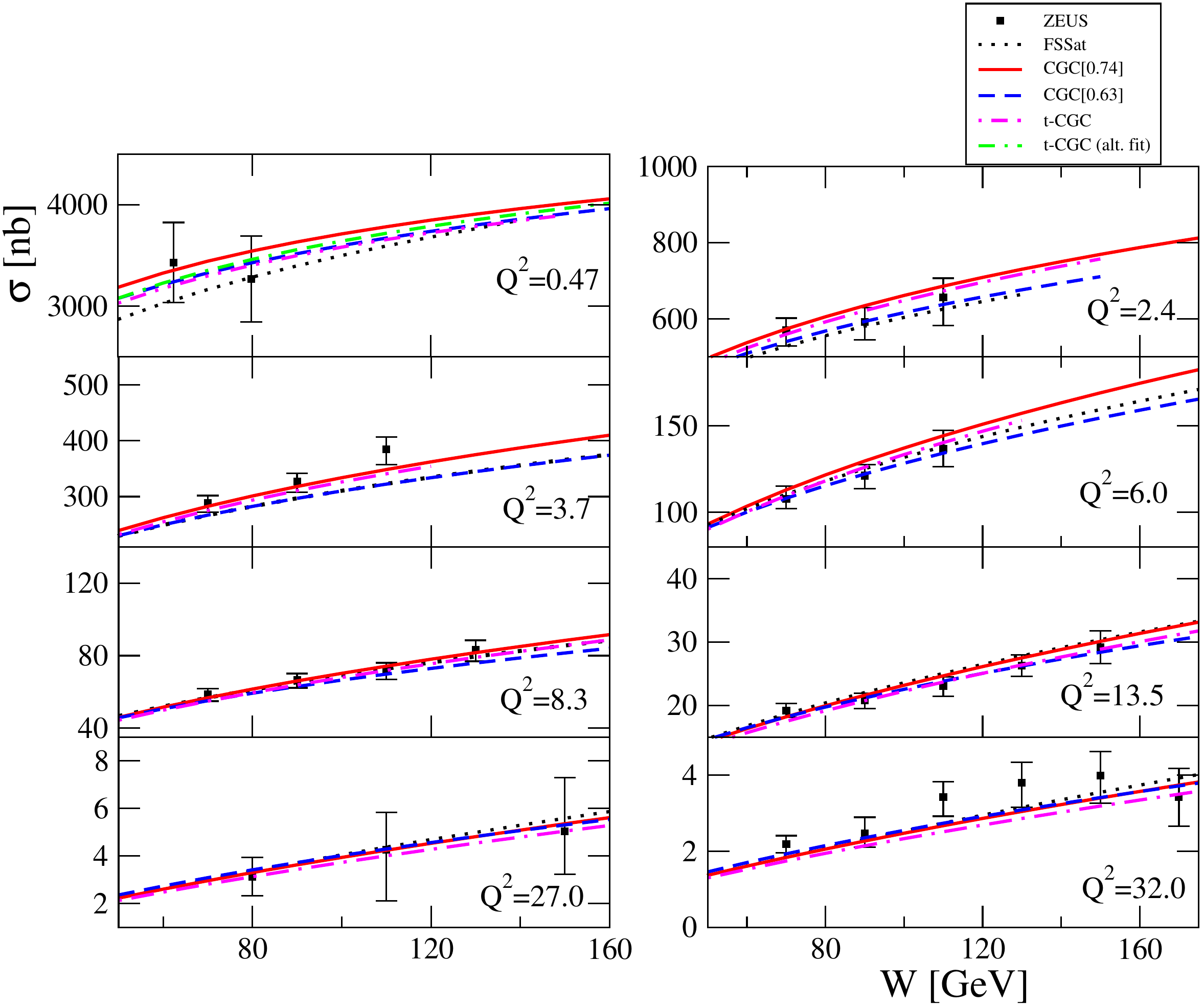}
\caption{Fits to the ZEUS total cross-section. Solid: CGC[0.74], dotted:
FSSat, dashed: CGC[0.63], dot-dashed: t-CGC. The alternative t-CGC fit is only
shown at $Q^2=0.47~\mbox{GeV}^2$ since at higher $Q^2$ it is indistinguishable
from the t-CGC fit.} 
\label{fig:ZEUS-xsec}
\end{figure}

\begin{figure}
\centering
\includegraphics*[width=9.cm]{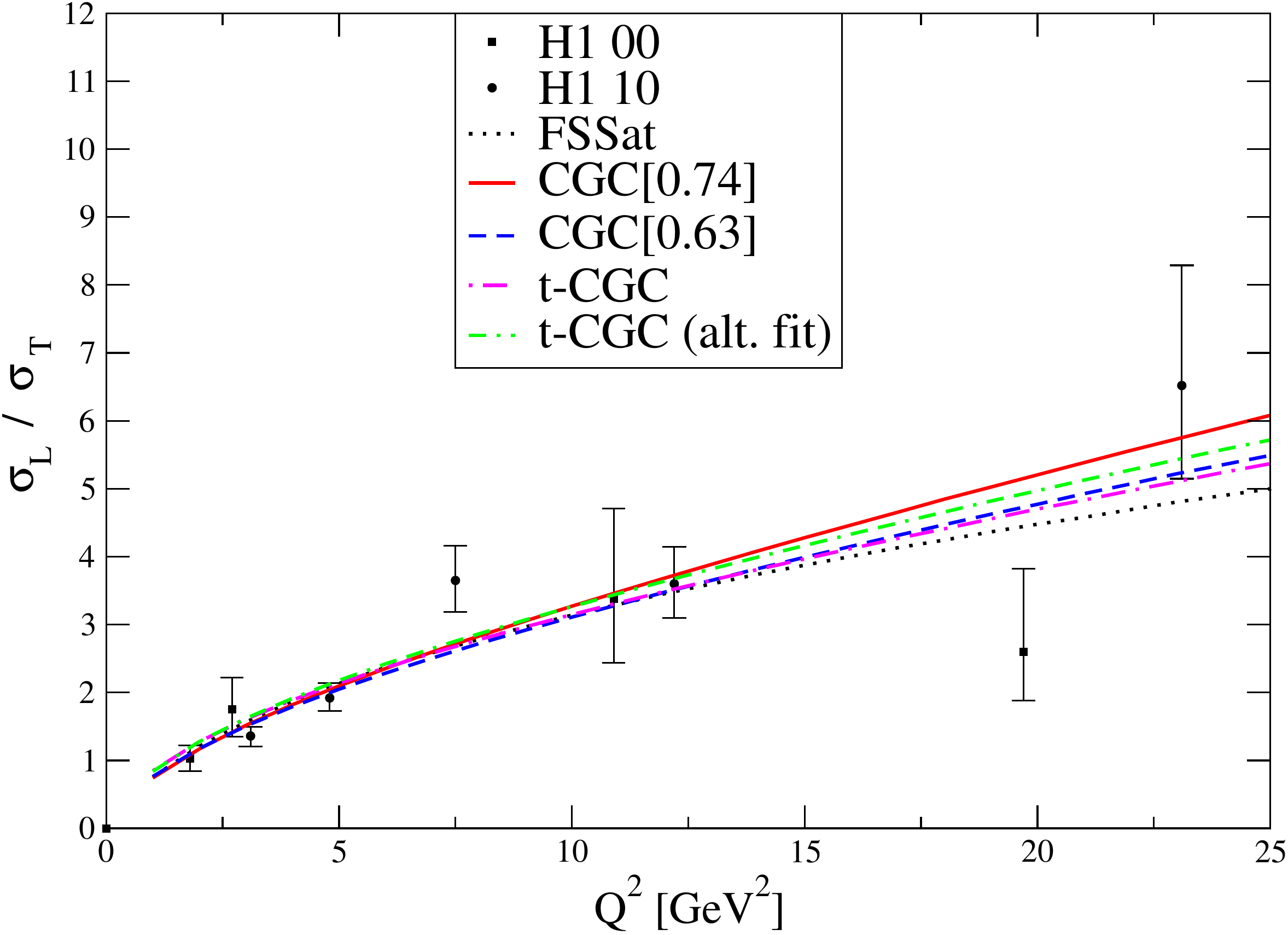}
\includegraphics*[width=9.cm]{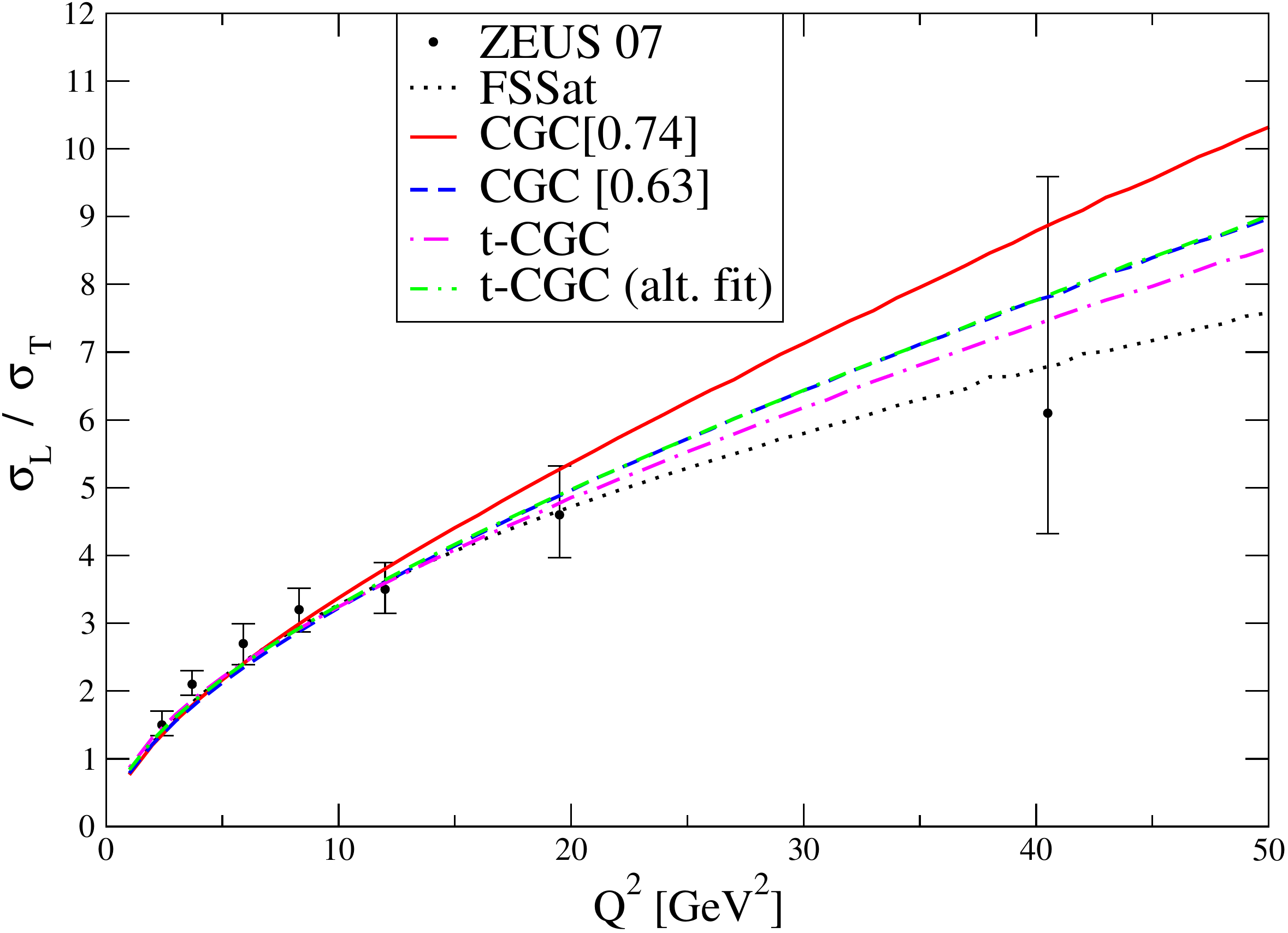}  
\caption{The $\sigma_L/\sigma_T$ data
are at $W = 75$~GeV for H1 and $W=90$~GeV for ZEUS. Solid: CGC[0.74],
dotted: FSSat, dashed: CGC[0.63], dot-dashed: t-CGC, dash-dash-dotted:
t-CGC(alt.).}
\label{fig:HERA-ratio}
\end{figure}

\begin{figure}\centering
\includegraphics*[width=15.cm]{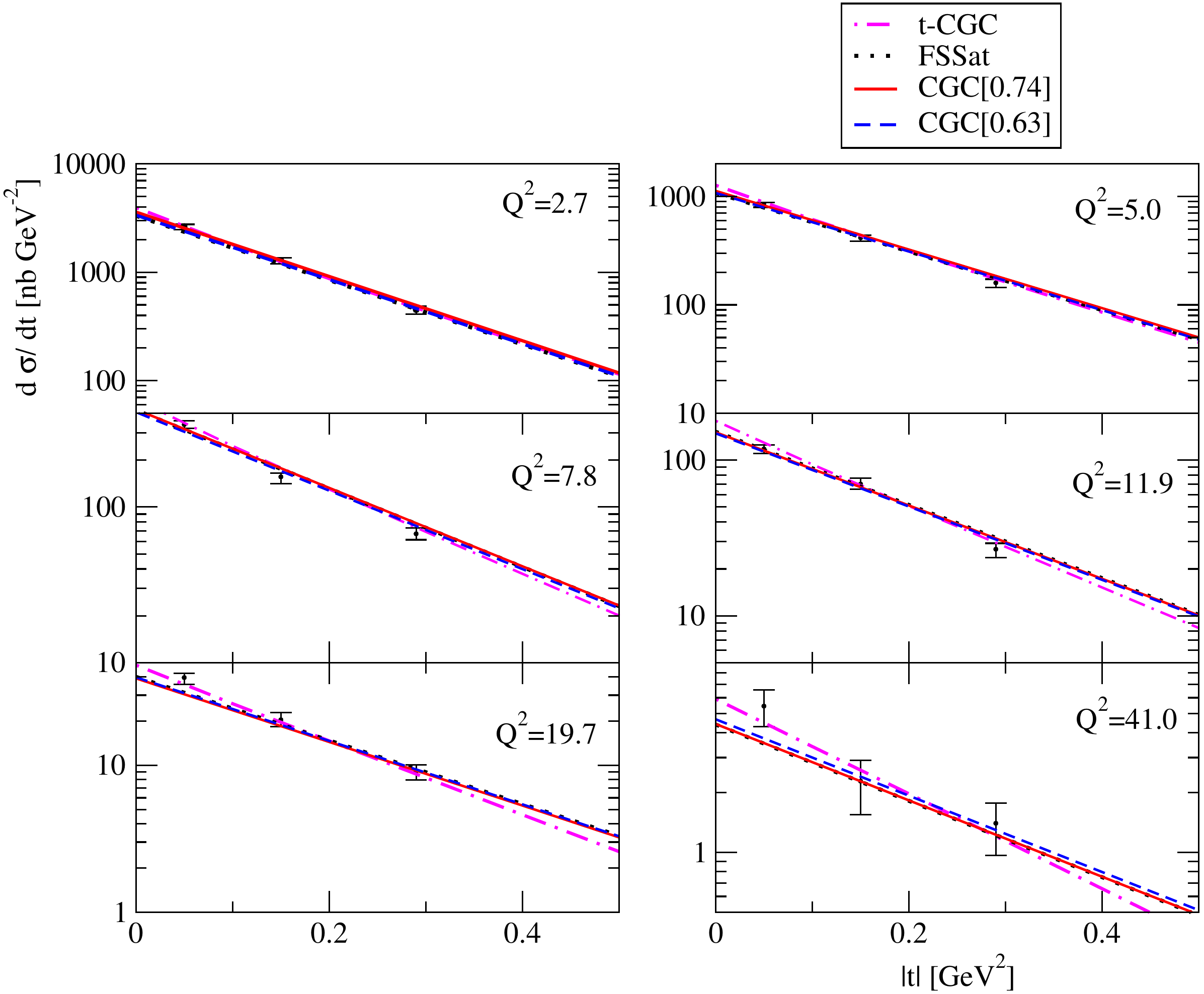}
\caption{Dot-dashed: t-CGC fit to the differential cross-section data from ZEUS
in the
range $0 \le |t| \le 0.5~\mbox{GeV}^2$. At these values of $Q^2$ the
alternative t-CGC(alt.) fit is indistinguishable from the t-CGC fit and is not
shown here. Also shown are the
predictions of the
forward models: Solid: CGC[0.74],
dotted: FSSat, dashed: CGC[0.63].}
\label{fig:ZEUS-dsigdt}
\end{figure}

\begin{figure}\centering
\includegraphics*[width=15.cm]{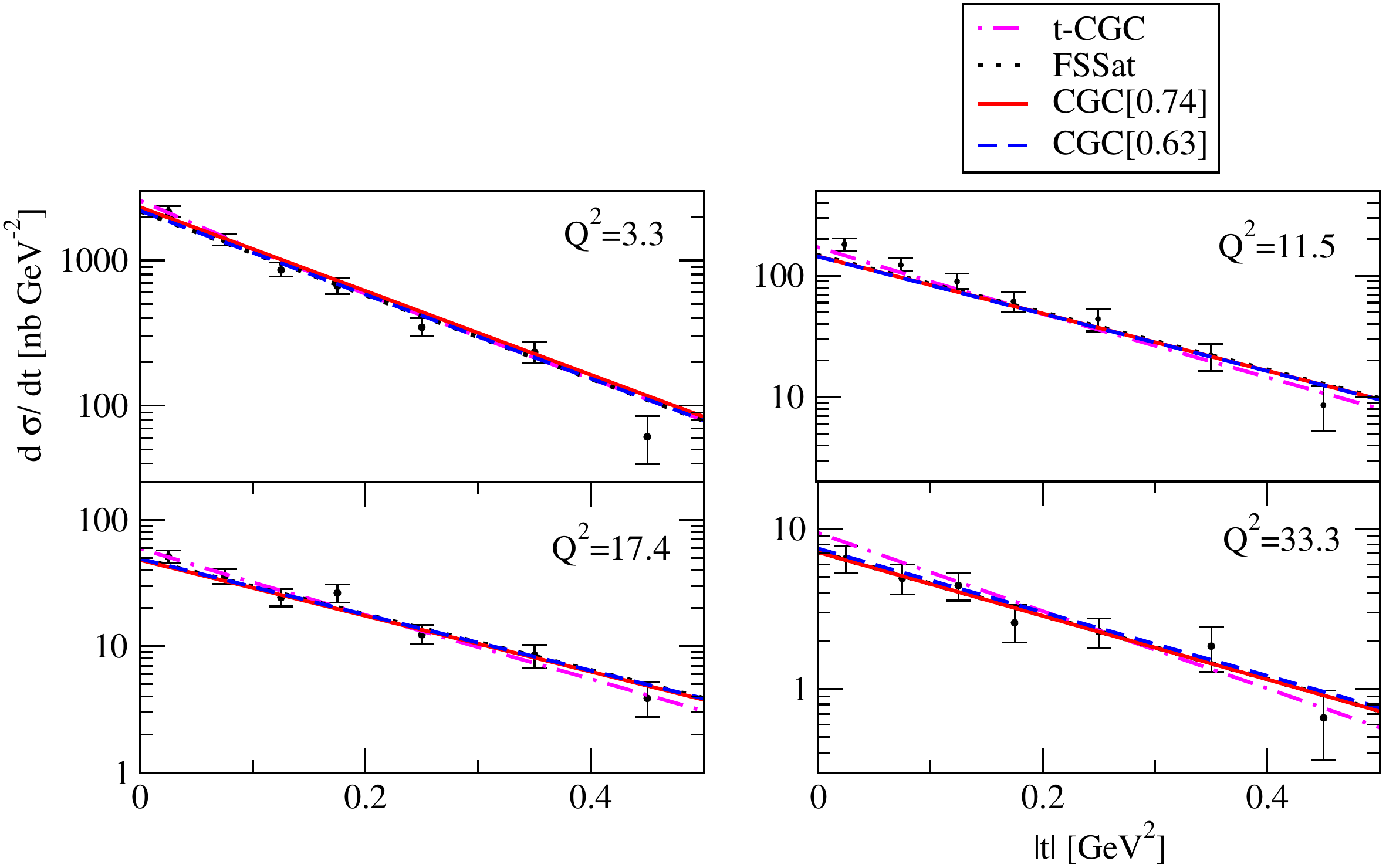}
\caption{Dot-dashed: t-CGC fit to the differential cross-section data from H1 in
the
range $0 \le |t| \le 0.5~\mbox{GeV}^2$. At these values of $Q^2$ the
alternative t-CGC(alt.) fit is indistinguishable from the t-CGC fit and is not
shown here. Also shown are the predictions of the
forward models: Solid: CGC[0.74],
dotted: FSSat, dashed: CGC[0.63].}
\label{fig:H1-dsigdt}
\end{figure}

Our fits to the HERA data are shown in figures
\ref{fig:H1-xsec} and \ref{fig:ZEUS-xsec} for the total cross-section, in figure
\ref{fig:HERA-ratio} for the $\sigma_L/\sigma_T$ ratio. In figures
\ref{fig:ZEUS-dsigdt} and \ref{fig:H1-dsigdt} we show the t-CGC fit to the
differential cross-section data as well as the predictions of the forward
models FSSat, CGC[$0.63$] and CGC[$0.74$].

\begin{table}[h]
\begin{center}
\textbf{Best fits for the meson wavefunction}
\[
\begin{array} 
[c]{|c|c|c|c|c|c|c|c|c|}\hline
\mbox{Reference} &\mbox{Model} & R_L^2 & R_T^2 & b_L & b_T & c_T
& d_T& \chi^2/\mbox{d.o.f} \\ \hline
\mbox{\cite{Forshaw:2010py,Boer:2011fh}}&\mbox{FSSat}     &26.76  &27.52
&0.5665 &0.7468 &0.3317 &1.310 &68/70  \\ \hline
\mbox{\cite{Boer:2011fh}}&\mbox{CGC}[0.63] &27.31  &31.92 &0.5522 &0.7289
&1.6927 &2.1457 & 67/70\\ \hline
\mbox{\cite{Boer:2011fh}}&\mbox{CGC}[0.74] &26.67  &21.30 &0.5697 &0.7929&0&0
& 64/72\\ \hline
\mbox{This paper}&\mbox{t-CGC} &29.63  & 21.64 &0.5018 &0.7396&0&0
&114/116 (72/75)\\ \hline
\mbox{This paper}&\mbox{t-CGC (alt.)} &29.68  & 20.96 &0.5004
&0.7314&-0.1601&-0.1656 &112/114 (70/75)\\ \hline
\end{array}
\]
\end{center}

\caption {Best fit wavefunction parameters and resulting $\chi^2/\mbox{d.o.f}$
obtained with each dipole model. For the
non-forward t-CGC model, the partial $\chi^2/\mbox{d.p}$ for the total
cross-section and
ratio data is given in brackets.}
\label{tab:best-fit-params}
\end{table}

\begin{figure}
\centering
\includegraphics*[width=10.cm]{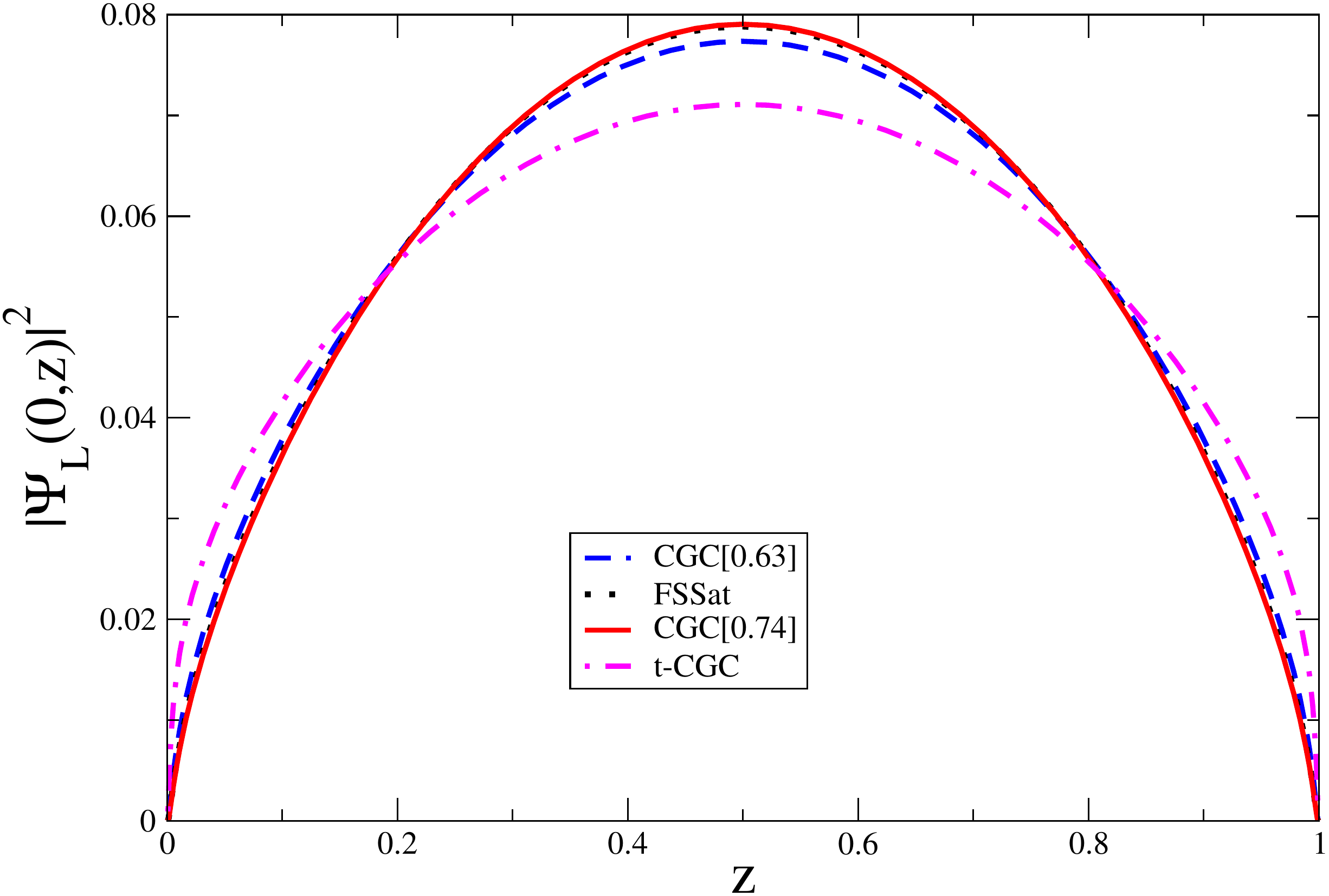}
\caption{The longitudinal light-cone wavefunction
squared evaluated at
$r=0$. Solid: CGC[$0.74$], Dotted: FSSat, Dashed: CGC[$0.63$], Dot-dashed:
t-CGC. The alternative t-CGC(alt.) fit is indistinguishable from the t-CGC fit.}
\label{fig:LCWF-L-r0}
\end{figure}

\begin{figure}
\centering
\includegraphics*[width=10.cm]{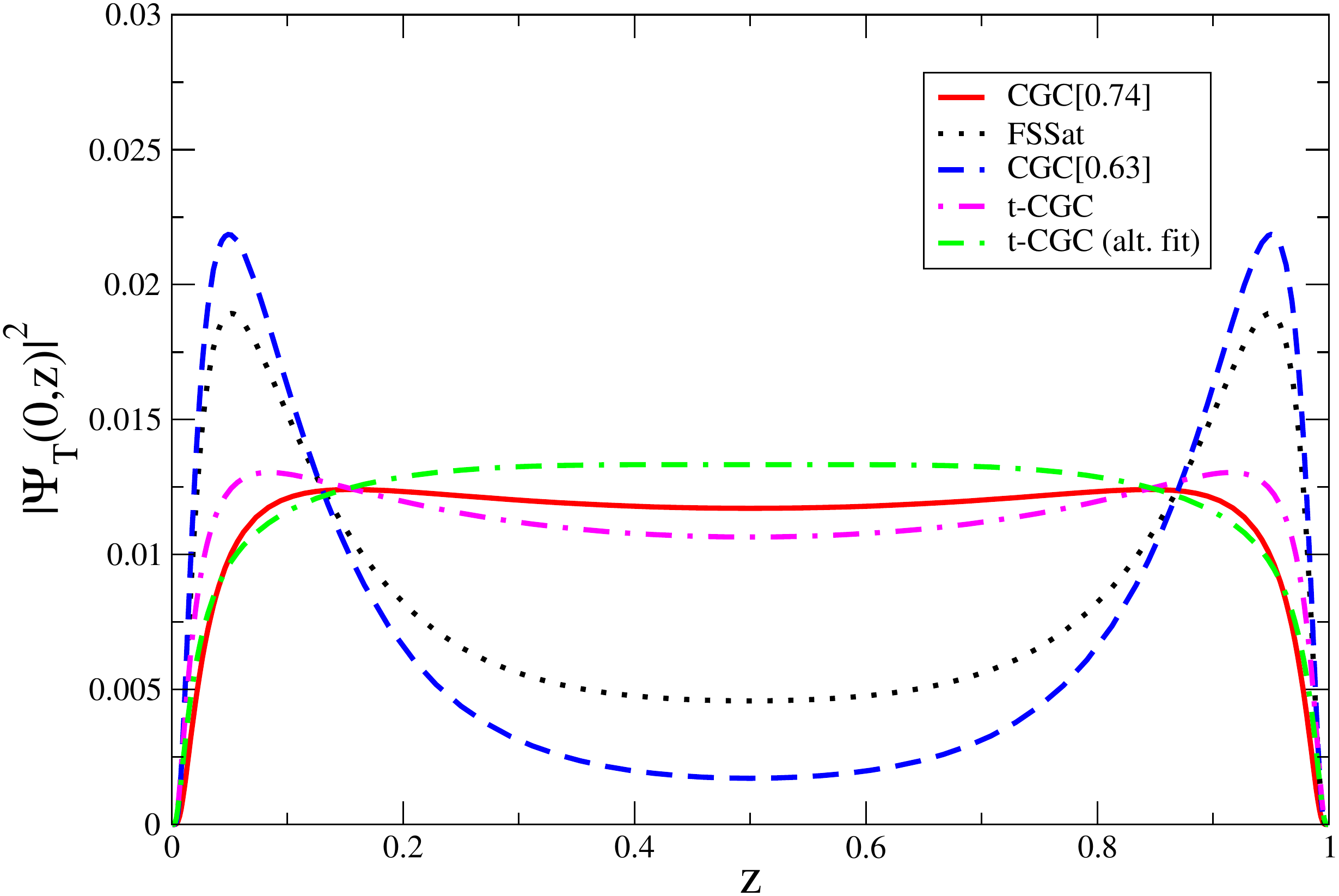}
\caption{The transverse light-cone wavefunction
squared evaluated at
$r=0$. Solid: CGC[$0.74$], Dotted: FSSat, Dashed: CGC[$0.63$], Dot-dashed:
t-CGC, dash-dash-dotted: t-CGC(alt.).}
\label{fig:LCWF-T-r0}
\end{figure}

In figures \ref{fig:LCWF-L-r0} and \ref{fig:LCWF-T-r0}, we show the extracted
light-cone wavefunctions squared, as defined by equation 
\eqref{Psisq-summed}, and evaluated at $r=0$. Figure \ref{fig:LCWF-L-r0}
confirms that the longitudinal wavefunction is not much model-dependent. On the
other hand, figure \ref{fig:LCWF-T-r0} shows that the data allow for a wide
range of end-point enhancement in the transverse wavefunction.

Additional theoretical constraints on the $\rho$ light-cone wavefunction can
be obtained from QCD Sum Rules and lattice QCD. More precisely, these
non-perturbative methods can predict the moments of the DA at a low
non-perturbative scale. In order to
use these constraints we must first establish the relationship between the DA
and the light-cone wavefunction.  We shall
then be in a position to compare the DA predicted by each of our extracted
wavefunctions with QCD Sum Rules and lattice predictions.

\section{Distribution Amplitudes}
Light-cone DAs appear in the operator product expansion of vacuum-to-meson
transition matrix
elements of quark-antiquark non-local gauge invariant operators at light-like
separations:\footnote{We use the
classification and notation of Ball and Braun~\cite{Ball:1996tb,Ball:1998fj}
with the notational simplication $g_{\perp}^{v} \rightarrow g_{\perp}$ since we
consider only a
vector coupling here. In the
alternative notation of
reference~\cite{Anikin:2009bf}, $\phi_\parallel=\varphi_1$ and
$g_{\perp}=\varphi_3$. We also note that in
references~\cite{Ball:1996tb,Ball:1998fj} the decay constant is defined as
$\sqrt{2} f_{\rho}$.} \cite{Ball:1998ff}
\begin{eqnarray}
\langle 0|\bar q(0) [0,x] \gamma^\mu q(x)|\rho (P,\lambda)\rangle 
&=& \sqrt{2} f_\rho M_\rho 
\frac{e_{\lambda}\cdot x}{P\cdot x}\, P^\mu \int_0^1 \d u \,e^{-iu P\cdot x}
\phi_\parallel(u,\mu)
\nonumber\\
&&{}+\left(e_{\lambda}^\mu-P^\mu\frac{e_{\lambda} \cdot x}{P \cdot x}\right)
\sqrt{2} f_{\rho}
M_{\rho}
\int_0^1 \d u\, e^{-iu P\cdot x} \,g_\perp(u,\mu) \nonumber\\
&&{}+ \{\cdots\}
\label{eq:DA-BB}
\end{eqnarray}
where $\mu$ is the renormalization scale, $x\cdot x=0$.  
The gauge link
\begin{equation}
[0,x] =\mbox{\rm Pexp}[ig\!\!\int_0^1\!\! \d t\,(-x)_\mu A^\mu((1-t)x)],
\label{Pexp}
\end{equation}
ensures the gauge invariance of the matrix elements. According to the
classification of Ball and Braun~\cite{Ball:1998fj,Ball:1998ff},
$\phi_{\parallel}$ is twist-$2$ and $g_{\perp}$ is twist-$3$. The $\{\cdots\}$
in equation \eqref{eq:DA-BB} stand for higher twist
contributions~\cite{Ball:1998fj,Ball:1998ff} which we do not consider here.

The $\mu$-dependent DAs 
\begin{equation}
\varphi=\{\phi_{\parallel},g_{\perp}\}
\end{equation}
satisfy the following normalisation condition:
\begin{equation}
 \int_0^1 \d u \; \varphi(u,\mu) = 1 \;.
\label{NormDA}
\end{equation}
This guarantees that in the local limit $x\rightarrow 0$, \eqref{eq:DA-BB}
becomes
\begin{equation}
\langle 0|\bar q(0) \gamma^\mu q(0)|\rho(P,\lambda)\rangle = \sqrt{2} f_{\rho}
M_{\rho} e^{\mu}_{\lambda}
\label{decay-constant-def}
\end{equation}
thereby recovering the definition of the decay constant $f_\rho$.

In order to establish a connection between these DAs and the light-cone
wavefunctions of the $\rho$, we need to apply the equal light-cone time
condition, $x^+=0$ and choose the light-cone gauge $A^+=0$.\footnote{It is
under these conditions that the Fock expansion of the physical $\rho$ is
carried out~\cite{Lepage:1980fj}.}  We can therefore
rewrite equation \eqref{eq:DA-BB} as
\begin{eqnarray}
\langle 0|\bar q(0)  \gamma^\mu q(0,x^-,0_{\perp})|\rho
(P,\lambda)\rangle 
&=& \sqrt{2} f_\rho M_\rho 
\frac{e_{\lambda} \cdot x}{P^+x^-}\, P^\mu \int_0^1 \d u \; e^{-iu P^+x^-}
\phi_\parallel(u,\mu)
\nonumber \\
&\hspace{-1.0cm}+&\hspace{-0.5cm} \sqrt{2} f_\rho M_\rho
\left(e_{\lambda}^\mu-P^\mu\frac{e_{\lambda} \cdot
x}{P^+x^-}\right)
\int_0^1 \d u \; e^{-iu P^+x^-} g_\perp(u,\mu)  
\label{eq:DA-BB-x0}
\end{eqnarray}
To isolate $\phi_{\parallel}$, we take $\mu=+$ with $\lambda=L$ in equation
\eqref{eq:DA-BB-x0} and obtain
\begin{equation}
\langle 0|\bar q(0) \gamma^+ q(x^-)|\rho
(P,L) \rangle = \sqrt{2} f_{\rho} P^+ \int \d u \; e^{-i u P^+ x^-} 
\phi_{\parallel}(u,\mu) \;.
\label{Long2}
\end{equation}
On the other hand, to isolate $g_{\perp}$, we take the scalar product of
equation \eqref{eq:DA-BB-x0} with the complex conjugate of the
meson's transverse polarization vector to arrive at
\begin{equation}
\langle 0|\bar q(0) e_T^* \cdot \gamma q(x^-)|\rho
(P,T) \rangle =  \sqrt{2} f_{\rho}M_{\rho} \int \d u \; e^{-i u P^+ x^-} 
g_{\perp}(u,\mu) \;.
\label{Trans2}
\end{equation}
Taking the Fourier transform of the above matrix elements with respect
to the longitudinal light-cone
distance $x^-$ yields: 
\begin{equation}
\phi_{\parallel}(z,\mu)= \frac{1}{\sqrt{2} f_{\rho}}\int \d x^- e^{i zP^+
x^-}\langle 0|\bar q(0)
\gamma^+ q(x^-)|\rho
(P,L) \rangle 
\label{Long4}
\end{equation}
and
\begin{equation}
g_{\perp}(z,\mu)= \frac{P^+}{\sqrt{2} f_{\rho}M_{\rho}}\int \d x^- e^{i zP^+
x^-}
\langle
0|\bar q(0)
e^{T*}.\gamma q(x^-)|\rho
(P,T) \rangle \;.
\label{Trans3}
\end{equation}

We now wish to express the matrix elements on the right-hand-sides of
the above equations \eqref{Long4} and \eqref{Trans3} in terms of the meson's
light-cone wavefunction given by equation \eqref{spinor-scalar-k}. To do so, we
use the Fock expansion
\begin{equation}
|\rho(P,\lambda) \rangle = \sqrt{4\pi N_c} \sum_{h,\bar{h}} \int \frac{
\d k^+ \d^2
\mathbf{k}}{16\pi^3\sqrt{k^+(P^+-k^+)}}  \Psi^{\rho,\lambda}_{h,\bar{h}}
(k^+/P^+,\mathbf{k})
\hat{b}_{h}^{\dagger}(k^+,\mathbf{k}) 
\hat{d}_{\bar{h}}^{\dagger}((P^+-k^+),-\mathbf{k}) |0\rangle 
\label{Fock-expansion-zero-Pt}
\end{equation}
and the mode expansion 
\begin{equation}
 q (0,x^-,\mathbf{0})= \int \left[\frac{\d k^ + \d^2\mathbf{k}}{16\pi^3
k^+}
\right] 
\sum_\alpha [\hat{b}_{\alpha} (k^+,\mathbf{k}) u_\alpha(k^+,\mathbf{k})
e^{-ik^+
x^-} + 
\hat{d}^{\dagger}_{\alpha} (k^+,\mathbf{k}) v_{\alpha} (k^+,\mathbf{k})
e^{ik^+
x^-}] 
\label{quark-field-expansion-lc}
\end{equation}
together with the equal light-cone time anticommutation relations
\begin{equation}
\{\hat{b}^{\dagger}_{h} (k^+,\mathbf{k}), \hat{b}_{h^{\prime}}
(k^{+\prime},\mathbf{k}^{\prime}) \}_{x^+=0}= 16\pi^3 k^+ \delta(k^+
-k^{+\prime})
\delta^2(\mathbf{k}-\mathbf{k}^{\prime}) \delta_{h h^{\prime}} ~\mbox{etc},
\label{anticommutation}
\end{equation}
in order to express the meson-to-vacuum matrix elements as
\begin{eqnarray}
 \langle 0 | \bar{q}(0) \gamma^{\mu} q(x^-) |\rho(P,\lambda)
\rangle &=&
\sqrt{4\pi N_c} \sum_{h,\bar{h}} \int \left[\frac{\d k^{+}\d^2\mathbf{k}
\Theta(|\mathbf{k}| <
\mu)}
{16\pi^3\sqrt{k^+(P^{+}-k^{+})}} \right]
\Psi^{\rho,\lambda}_{h,\bar{h}}(k^+/P^+,\mathbf{k})\nonumber \\
&\times& 
\bar{v}_{\bar{h}}(P^{+}-k^{+},-\mathbf{k}) \gamma^{\mu}
u_h(k^+,\mathbf{k})
e^{-ik^{+}
x^{-}} \;.
\label{evaluated-matrix-element}
\end{eqnarray}
Here we have identified the renormalization scale $\mu$ with the ultraviolet
cut-off on
transverse momenta~\cite{Kogut:1973ub,Diehl:2002he}.

Taking the Fourier transform with respect to the light-cone distance $x^-$
allows us to carry out the integral over $k^+$, which in turn fixes the
plus-momentum
of the quark to be $zP^+$ and so
\begin{eqnarray}
\label{evaluated-matrix-element-delta}
P^+\int \d x^- e^{ix^-zP^+} \langle 0 | \bar{q}(0)  \gamma^{\mu}
q(x^-)
|\rho(P,\lambda) \rangle &=& \sqrt{4\pi N_c} \sum_{h,\bar{h}}
\int^{|\mathbf{k}| < \mu}
\frac{\d^2\mathbf{k}}{16\pi^3}
\Psi^{\rho,\lambda}_{h,\bar{h}}(z,\mathbf{k})\\
\nonumber 
&\times&
\left \{ \frac{\bar{v}_{\bar{h}}((1-z)P^{+},-\mathbf{k})}{\sqrt{(1-z)}}
\gamma^{\mu} \frac{u_h(zP^+,\mathbf{k})}{\sqrt{z}} \right \}
\;.
\end{eqnarray}
We can now take the scalar product with the
meson's transverse polarisation vector to obtain
\begin{equation}
P^+ \int \d x^- e^{ix^-zP^+} \langle 0 | \bar{q}(0) e^{T
*}.\gamma
q(x^-)
|\rho(P,T) \rangle = N_c
\sum_{h,\bar{h}} \int^{|\mathbf{k}| < \mu} \frac{\d^2\mathbf{k}}{16\pi^3}
|S_{h,\bar{h}}^{\rho,T}(z,\mathbf{k})|^2 \phi_{T}(z,\mathbf{k}) 
\label{evaluated-matrix-element-photon-T}
\end{equation}
and with the help of equation \eqref{Trans3}, we can deduce that
\begin{equation}
g_{\perp}(z,\mu) =
\frac{N_c}{\sqrt{2} f_\rho M_\rho}
\sum_{h,\bar{h}} \int^{|\mathbf{k}| < \mu} \frac{\d^2\mathbf{k}}{16\pi^3}
|S_{h,\bar{h}}^{\rho,T}(z,\mathbf{k})|^2 \phi_{T}(z,\mathbf{k}) \;.
\label{evaluated-matrix-element-photon-T-g}
\end{equation}
On the other hand, if we consider $\lambda=L$ and $\mu=+$, equation
\eqref{evaluated-matrix-element-delta} becomes
\begin{equation}
\int dx^- e^{ix^-zP^+} \langle 0 | \bar{q}(0) \gamma^+
q(x^-)
|\rho(P,L) \rangle = 
-2 N_c \sum_{h,\bar{h}} \int^{|\mathbf{k}| < \mu} \frac{\d^2\mathbf{k}}{16\pi^3}
S_{h,\bar{h}}^{\rho,L}(z,\mathbf{k}) \phi_{L}(z,\mathbf{k}) \delta_{h,-\bar{h}}
\label{evaluated-matrix-element-photon-L}
\end{equation}
and so
\begin{equation}
\phi_\parallel (z,\mu) =-\frac{2 N_c}{\sqrt{2} f_\rho}
\sum_{h,\bar{h}} \int^{|\mathbf{k}| < \mu} \frac{\d^2\mathbf{k}}{16\pi^3}
S_{h,\bar{h}}^{\rho,L}(z,\mathbf{k}) \phi_{L}(z,\mathbf{k}) \delta_{h,-\bar{h}}
\;.
\label{evaluated-matrix-element-photon-L-phi}
\end{equation}
Using equations \eqref{SL} and \eqref{ST}, we can rewrite
\eqref{evaluated-matrix-element-photon-T-g} and
\eqref{evaluated-matrix-element-photon-L-phi} as
\begin{equation}
 g_\perp(z,\mu)=\frac{N_c}{2 \pi \sqrt{2} f_{\rho} M_{\rho}} \int^{|\mathbf{k}|
<
\mu}
\frac{\d^2\mathbf{k}}
{(2\pi)^2} [m_f^2 + (z^2 + (1-z)^2) \mathbf{k}^2]
\frac{\phi_T(z,\mathbf{k})}{z^2(1-z)^2} 
\label{DA-scalar-lcwf-k-space-T}
\end{equation}
and
\begin{equation}
\phi_{\parallel}(z,\mu) =\frac{N_c}{ \pi M_{\rho}\sqrt{2} f_{\rho}}
\int^{|\mathbf{k}| < \mu}
\frac{\d^2\mathbf{k}}{(2\pi)^2}  [M_\rho^2 z (1-z)
+ m_f^2 + \mathbf{k}^2] \frac{\phi_L(z,\mathbf{k})}{z(1-z)} 
\label{DA-scalar-lcwf-k-space-L}
\end{equation}
respectively.

Finally, inserting the Fourier transform pairs
\begin{equation}
\int \d^2\mathbf{r} \; e^{-i \mathbf{k} \cdot \mathbf{r}} \phi_{\lambda}
(z,\mathbf{r})=
\phi_{\lambda} (z,\mathbf{k})
\end{equation}
and 
\begin{equation}
 \int \d^2\mathbf{r} e^{-i \mathbf{k} \cdot \mathbf{r}} \{-\nabla^2
\phi_{\lambda}
(z,\mathbf{r}) \}
= \mathbf{k}^2
\phi_{\lambda} (z,\mathbf{k})
\end{equation}
in \eqref{DA-scalar-lcwf-k-space-T} and \eqref{DA-scalar-lcwf-k-space-L} and 
carrying out the integration over $|\mathbf{k}|$, we arrive at
\begin{equation}
\phi_\parallel(z,\mu) =\frac{N_c}{\pi \sqrt{2} f_{\rho} M_{\rho}} \int \d
r \mu
J_1(\mu r) [M_{\rho}^2 z(1-z) + m_f^2 -\nabla_r^2] \frac{\phi_L(r,z)}{z(1-z)}
\label{DA-scalar-lcwf-r-space-L}
\end{equation}
and 
\begin{equation}
g_\perp(z,\mu)=\frac{N_c}{2 \pi \sqrt{2} f_{\rho} M_{\rho}} \int \d r \mu
J_1(\mu r)
\left[ (m_f^2 - (z^2+(1-z)^2) \nabla_r^2 \right] \frac{\phi_T(r,z)}{z^2 (1-z)^2
}
\;.
\label{DA-scalar-lcwf-r-space-T} 
\end{equation}
Equations \eqref{DA-scalar-lcwf-r-space-L} and
\eqref{DA-scalar-lcwf-r-space-T} are
our main results. They show that the DAs can be expressed in terms of the
scalar parts of the light-cone wavefunctions.  

Note that in the limit $\mu \rightarrow \infty$, equation
\eqref{DA-scalar-lcwf-r-space-L} becomes
\begin{equation}
\phi_\parallel(z,\infty) =\frac{N_c}{\pi \sqrt{2} f_{\rho} M_{\rho}}
\left. [M_{\rho}^2 z(1-z) + m_f^2 -\nabla_r^2]
\frac{\phi_L(r,z)}{z(1-z)}\right|_{r=0}
\label{DA-scalar-lcwf-r-space-L-mu-infty}
\end{equation}
so that the decay width constraint given by equation \eqref{longdecay} implies
that our extracted twist-$2$ DA $\phi_{\parallel}$ satisfies the normalisation
condition given by equation \eqref{NormDA}. On the other hand, our extracted
twist-$3$ DA $g_{\perp}$ satisfies only approximately this normalisation
condition.

\section{Comparison with Sum Rules and lattice predictions}
To make contact with Sum Rules predictions, we expand the
twist-$2$ DA in Gegenbauer polynomials, i.e.\cite{Ball:1998ff}
\begin{equation}
\phi_{||}(z,\mu)=6 z(1-z) \left [1 + \sum_{j=2,4,...} a_{j}^{||} (\mu)
C_{j}^{3/2}(\xi) \right] \;,
\label{gegenbauer_expansion}
\end{equation}
where $C_{j}^{3/2}$ are the Gegenbauer polynomials. The factor
$z(1-z)$ is the weight function over which the Gegenbauer polynomials are
orthogonal. Keeping only
the lowest conformal spin contribution in the expansion yields
\begin{equation}
\phi_\parallel(z,\mu) =  6 z(1-z) \left[ 1+
a_2^\parallel (\mu) \, \frac{3}{2} ( 5\xi^2  - 1 ) \right] \;.
\label{eq:phipar}
\end{equation}
Similarly an explicit expression for the twist-$3$ DA
is\cite{Ball:1998ff}
\begin{eqnarray}
  g_\perp(z,\mu) & = & \frac{3}{4}(1+\xi^2)
 + \left(\frac{3}{7} \, 
a_2^\parallel(\mu) + 5 \zeta_{3}(\mu) \right) \left(3\xi^2-1\right)
 \nonumber\\
& & {}+ \left[ \frac{9}{112}\, a_2^\parallel(\mu) 
+ \frac{15}{64}\, \zeta_{3}(\mu) \Big(3\,\omega_{3}^V(\mu)-\omega_{3}^A
(\mu)\Big)
 \right] \left( 3 - 30 \xi^2 + 35\xi^4\right) \;.
\label{eq:gv}
\end{eqnarray}


The QCD sum rules estimates for the various non-perturbative
parameters have been updated in reference \cite{Ball:2007zt}:
$a_{2}^{||}=0.15 \pm 0.07$,
$\zeta_3=0.030 \pm 0.010$, $\omega_3^V=5.0\pm 2.4$ and
$\omega_3^A=-3.0 \pm 1.4$, at a scale of $\mu=1$ GeV. Their perturbative
evolution with the scale $\mu$ are explicitly given in
reference~\cite{Ball:1998ff}. As $\mu
\rightarrow \infty$, all these parameters vanish and we obtain the
asymptotic DAs given by
\begin{equation}
 \phi_{\parallel} (z,\infty)=6z(1-z)
\label{asymp-DA-L}
\end{equation}
and
\begin{equation}
g_{\perp}(z,\infty)=\frac{3}{4}(1+\xi^2)
\;.
\label{asymp-DA-T}
\end{equation}

We are now able to compare our extracted DAs computed using equation
\eqref{DA-scalar-lcwf-r-space-L} and 
\eqref{DA-scalar-lcwf-r-space-T} with the
QCD Sum Rule prediction given by
equation \eqref{eq:phipar} and \eqref{eq:gv} respectively. We first confirm our
earlier
observation~\cite{Forshaw:2010py} that our
extracted DA hardly evolves with the scale $\mu$ when $\mu \ge 1$ GeV, i.e. it
neglects the perturbatively known $\mu$-dependence. Given the limited $Q^2$
range of the HERA data to which we fit ($\sqrt{Q^2} < 7~$ GeV), our extracted DA
should thus be viewed as a parametrization at some low scale $\mu \sim 1$ GeV.

\begin{figure}
\centering
\includegraphics*[width=10cm]{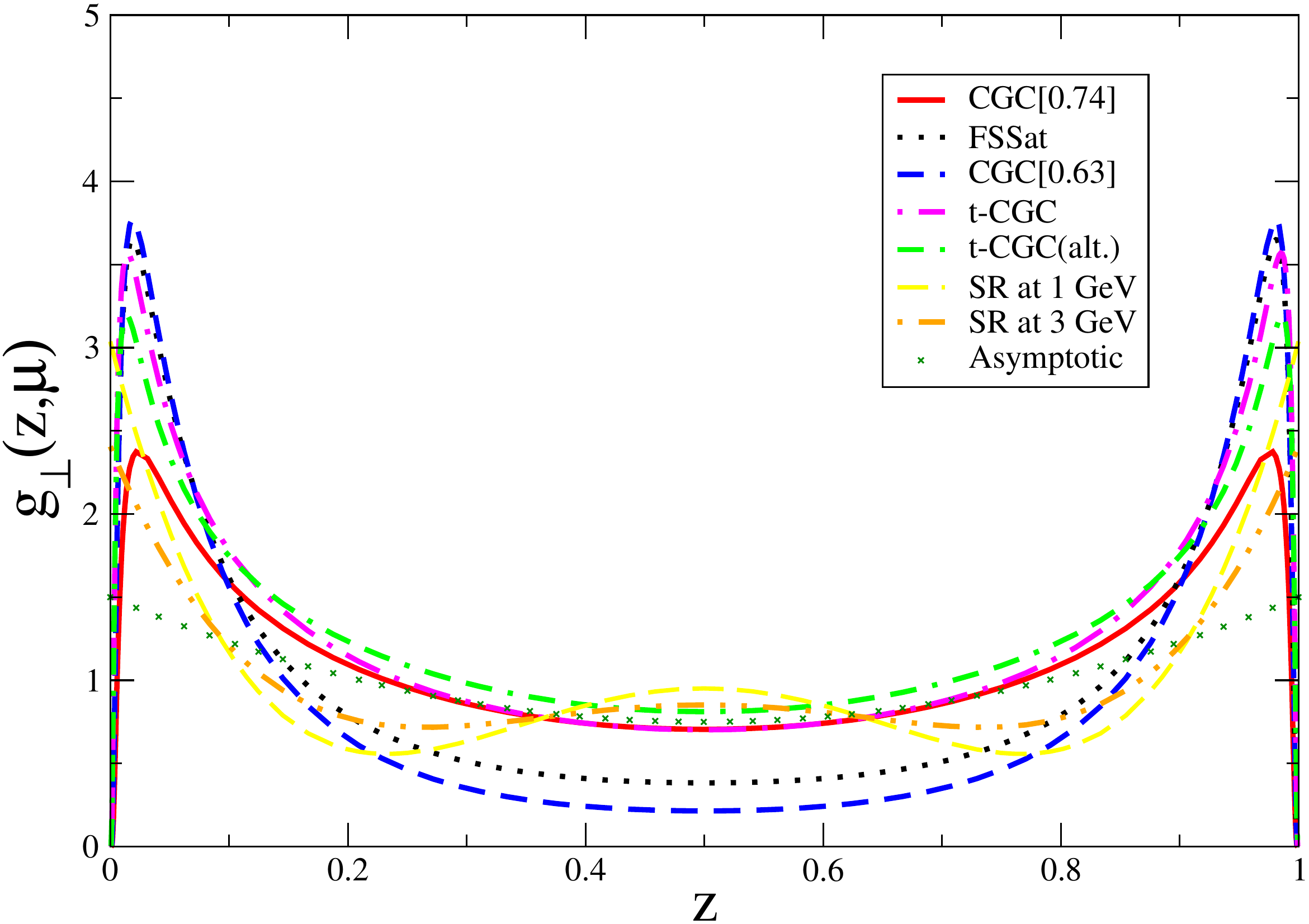}
\caption{The extracted twist-$3$ DAs compared to the QCD
Sum Rules predictions. Dotted: FSSat; Solid: CGC[$0.74$];  Dashed: CGC[$0.63$];
Dot-dashed: t-CGC; Dash-dash-dotted: t-CGC (alt.); Long-dashed: Sum Rules at $1$
GeV;
Dot-dot-dashed: Sum Rules at $3$ GeV. Crosses: Asymptotic.}
\label{fig:DAT}
\end{figure}

We compare in figure \ref{fig:DAT} our extracted twist-$3$ DAs with the Sum Rule
DA at two values of $\mu$: $1$ and $3$ GeV. Note that we obtain the DA at $3$
GeV by evolving the DA at $1$ GeV to leading logarithmic
accuracy~\cite{Ball:1998ff}. All the distributions show an enhanced end-point
compared to the asymptotic DA given by equation \eqref{asymp-DA-T} and which is
also shown on the plot.  Note that the non-monotonic behaviour of the Sum Rule
distribution is not physical and is due to the truncation in the
Gegenbauer expansion. 


\begin{figure}
\centering
\includegraphics*[width=10cm]{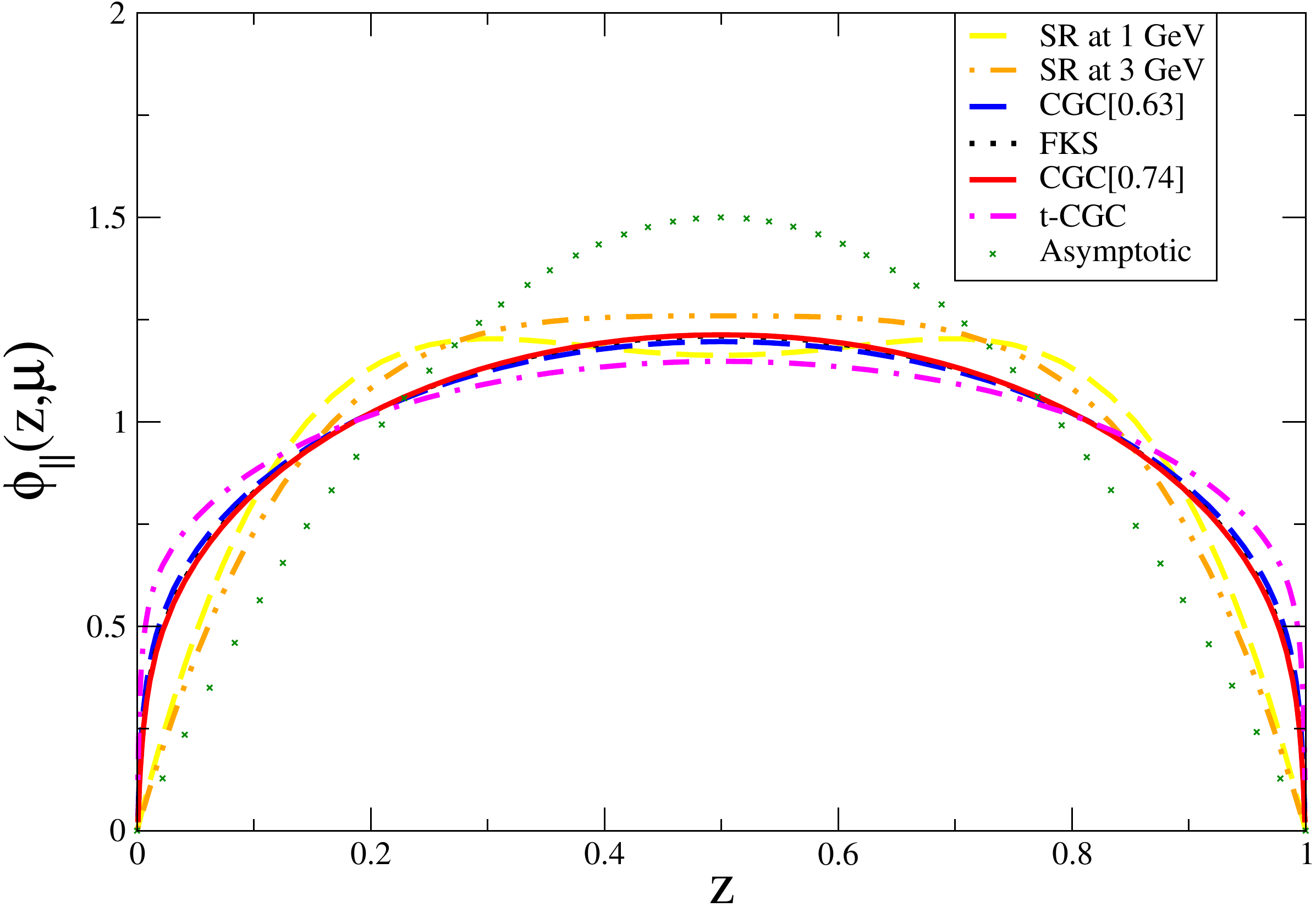}
\caption{The extracted leading twist-$2$ DAs compared to the
Sum Rule predictions. Dotted: FSSat; Solid: CGC[$0.74$];  Dashed: CGC[$0.63$];
Dot-dashed: t-CGC; Long-dashed: Sum Rules at $1$ GeV;
Dot-dot-dashed: Sum Rules at $3$ GeV. Crosses: Asymptotic.}
\label{fig:DAL}
\end{figure}

In figure \ref{fig:DAL}, we  show our extracted twist-$2$ DAs
which are sensitive to the longitudinal wavefunction, i.e. computed using
equation \eqref{DA-scalar-lcwf-r-space-L}. Similar predictions appeared
in
reference \cite{Forshaw:2010py,Boer:2011fh} for the FSSat and the forward CGC
models. We now
also show the twist-$2$ DA extracted with the non-forward t-CGC
model and in addition we use equation \eqref{DA-scalar-lcwf-r-space-L} to
compute the DA instead of assuming that the DA is simply proportional to the
scalar light-cone wavefunction integrated over transverse
momentum~\cite{Brodsky:1994kf}. 

The extracted DAs are all consistent with the
QCD Sum Rule prediction at $1$ GeV, given the size of the uncertainty on the
parameter $a_2^{\parallel}$. Table \ref{tab:moments-mu}
compares the corresponding second moments of the twist-$2$ DA given by
\begin{equation}
\langle \xi^2 \rangle_{\mu} = \int_0^1 \d z \; \xi^2 \phi_{\parallel}(z,\mu)
\;. 
\end{equation}
We also show in this table
the prediction of the old Boosted Gaussian wavefunction which does not fit the
HERA data.


\begin{table}[h]
\begin{center}
\textbf{Moments of the leading twist DA at the scale $\mu$}
\[
\begin{array} 
[c]{|c|c|c|c|}\hline
\mbox{Reference} & \mbox{Approach} & \mbox{Scale}~\mu &\langle \xi^2
\rangle_{\mu}\\
\hline
\mbox{\cite{Forshaw:2010py}} & \mbox{Old BG prediction}&\sim 1~\mbox{GeV}
&0.181\\ \hline
\mbox{This paper} & \mbox{FSSat fit}&\sim 1~\mbox{GeV}
&0.267\\ \hline
\mbox{This paper} & \mbox{CGC[$0.74$]}&\sim 1~\mbox{GeV} &0.266\\
\hline
\mbox{This paper} & \mbox{CGC[$0.63$]}&\sim 1~\mbox{GeV} &0.271 \\
\hline
\mbox{This paper} &\mbox{t-CGC}&\sim 1~\mbox{GeV} &0.286\\ \hline
\mbox{\cite{Boyle:2008nj}}&\mbox{Lattice} &2~\mbox{GeV} &0.24(4) \\
\hline 

\mbox{\cite{Ball:2007zt}}&\mbox{Sum Rules}&1~\mbox{GeV} &0.254\\ \hline
\mbox{\cite{Ball:2007zt}}&\mbox{Sum Rules}&3~\mbox{GeV} &0.237\\ \hline
& 6z(1-z)
&\infty&0.2 \\ \hline
\end{array}
\]
\end{center}
\caption {Our extracted values for  $\langle \xi^2 \rangle_{\mu}$, compared to
predictions based on the QCD Sum Rules or lattice QCD. Note that the Sum Rules
prediction at $\mu=1$ GeV has an uncertainty of $\pm 0.024$ corresponding to
the uncertainty of $0.07$ in $a_2^{\parallel}$.}
\label{tab:moments-mu}
\end{table}

\section{Conclusions}

We have extracted the leading twist-$2$ and subleading twist-$3$
Distribution Amplitudes of the $\rho$ meson using the HERA
data. We find that the twist-$2$ DA is not much model dependent 
where as the data
allow for a family of extracted twist-$3$ DAs
with a varying degree of end-point enhancement. The extracted DAs are consistent
with the QCD Sum Rules and lattice predictions.

In its present form, our parametrization for the DA lacks the perturbative QCD
evolution. Improving this would enable a more precise comparison with Sum Rules
and the lattice.

\section{Acknowledgements}
R.S. acknowledges the hospitality of the Particle Physics
Group of the University of Manchester where parts of this work were
carried out. We thank M. Diehl for
useful discussions. This research is supported
by the UK's STFC.

\bibliographystyle{JHEP}
\bibliography{LCDArho}

\end{document}